# Wedge disclination description of emergent core-shifted grain boundaries at free surfaces

**Authors:** Xiaopu Zhang,[*] Ipen Demirel, John J. Boland[*]

**Affiliations:**

Centre for Research on Adaptive Nanostructures and Nanodevices (CRANN), AMBER SFI Research Centre and School of Chemistry, Trinity College Dublin, Dublin 2, Ireland

*Correspondence to: xiaopuz@tcd.ie, jboland@tcd.ie

**Abstract**

Emergent grain boundaries at free surface control material properties such as nanomaterial strength, catalysis, and corrosion. Recently the restructuring of emergent boundaries on copper (111) surfaces was discovered experimentally and atomic calculations point to its universality in fcc metal systems. Restructuring is due to a preference for boundaries to shift their tilt axis across the $(1\bar{1}0)$ plane towards [112] and ultimately to form low energy [112] core shifted boundaries (CSBs). However, the observed geometry of these emergent boundaries is not reproduced by atomic calculations and the driving force is still controversial due to inconsistencies between the computational continuum analysis and atomic calculations. Here, using atomic calculations that involve a methodical shift of the dislocation core, we confirmed the core shift of emergent boundaries observed in experiment and reconciled the atomic calculations with the elastic analysis through the inclusion of a straight wedge disclination at the free surface.

## 1. Introduction

The properties of emergent grain boundaries (eGBs) at free surfaces have attracted attention for more than half a century [1]. Intensive research has explored the effect of eGBs on growth and residual stress evolution in thin films and coating [2-10], electrocatalytic phenomena such as $CO_2$ reduction [11, 12], chemical catalysis [13], intergranular corrosion [14-16], and the mechanical properties of nanocrystalline materials [17, 18]. Recently, the restructuring of emergent boundaries in nanocrystalline copper films and macroscopic bicrystal copper samples with (111) surfaces was discovered experimentally [19, 20]. Restructuring involves the formation of core shifted boundaries that is accompanied by the rotation of adjoining grains and the generation of elastic stresses in the triple junction region. The energetics of eGB formation was analyzed by comparing the energetics from continuum elastic modelling and atomic calculations, in attempt to reproduce the observed eGB structure. Computational continuum elastic analysis of the deformation of the film perpendicular to the boundary plane was carried out by considering the elastic anisotropy of the single crystalline material and elastic properties of textured polycrystals [19]. Only a small number of low-energy atomic structures were identified based on a cut-paste-shear-stitch method developed via a genetic algorithm and Monte-Carlo method [19]. However, the level of agreement and consistency between the elastic modelling and atomic calculations was limited.

In this paper, we introduce the core-shift method to systematically build restructured eGBs, instead of relying on a structure search. In this manner, it is possible to build all possible low energy core-shifted eGBs. We then analyzed the core structure of these calculated [112] core-shifted eGBs, uncovered the principle underlying our framework and demonstrated its wide applicability and consistency. Our analysis of the energetics of [112] core-shifted eGBs shows that core-shifted eGBs at free surfaces can be well described as a straight wedge disclination from the perspective of wedge volume, the elastic stress and the surface stress.

## 2. Calculation & results

Experiments showed that emergent grain boundaries (eGB) at surfaces of Cu (111) nanocrystalline films and bicrystals have tilt axes that are locally shifted away from [111] [19, 20].

This restructuring phenomenon involves the shifting of boundary cores across the $(1\bar{1}0)$ boundary plane and ultimately the formation of low energy [112] core shifted boundaries (CSBs), in which the cores then lie along the $[11\bar{1}]$ close packed plane. Core-shifting drives an out-of-plane rotation of the adjoining grains on either side of the boundary. The relationship between the in-plane boundary angle $\theta$, the out-of-plane rotation angle $\varphi$ and the core shift angle (or inclination angle) $\psi$ away from [111] has been established from the boundary geometry and verified by experiments. The predicted trend $\tan\psi = \tan(\varphi/2)/\sin(\theta/2)$ was found experimentally for low angle [111] symmetrical tilt grain boundaries [20, 21].

However, for grain boundaries with in-plane angles between ~ 15° to 30° shown schematically in Fig. 1a-b, the predicted out-of-plane rotation (Fig. 1c) was observed only locally close to the triple junction of the emerging boundary, while the global out-of-plane angle far from the boundary was close to zero [19]. This implies that while there is a shift of the tilt axis from [111] to [112] close to the surface, deep within the material the tilt axis is still [111] as shown in Fig. 1c-d. This complex boundary structure that involves a local symmetric rotation of the adjoining grains is most appropriately described as a straight wedge disclination (Fig. 1c). At present there is no unified self-consistent description of [112] CSB formation that accounts for the depth dependence of the associated disclination.

To develop a systematic approach to model the energetics and depth of [112] CSB formation, we first consider the emergence of a single dislocation onto a surface (Fig 1e). To shift the boundary core in each layer at the surface, deleting or inserting core atoms is necessary, as illustrated in Fig. 1e [22]. In the schematic shown, deleting one core atom shifts the dislocation line upward in the boundary plane while adding one atom below the core atoms shift the dislocation line downward. Both are essentially equivalent to diffusion controlled dislocation climb by jog nucleation and elimination [23, 24]. Core atom deletion (insertion) creates valley (ridge) emergent boundaries, as described previously [19]. Using this approach, we developed a methodology to systematically evaluate the energetics of the core-shifting process, that keeps track of the number of atoms that are deleted or inserted and the formation of the associated wedge disclination.

We carried out calculations with LAMMPS software using the embedded-atom-method (EAM) interatomic potentials for Cu [25, 26]. Boundaries in bulk are calculated using molecular statics (MS) with periodic boundary conditions in three directions and each repeat cell contains a pair of parallel GBs of equal and opposite misorientation [20, 27, 28]. The geometrical specification is listed in Tab. S1. Structure searches were carried out and only the lowest-energy structures are reported here [28].

The suspended films are built by stacking relaxed boundaries of ABC layers along [111] direction with the top layer C stacking, and adding an additional vacuum layer, as detailed in Fig. S1. The defect energy or the excess energy as a result of the presence of the interfaces, comprised of the top and bottom surfaces and two boundaries, and the four ideal eGBs [1, 29] is $E_N = E_N^0 + \varepsilon_i$, where the interfacial energy is $E_N^0 = 2\gamma_{gb} \cdot A_{gb} + 2\gamma_s \cdot A_s$, $\gamma_{gb}$ is the boundary energy, $\gamma_s$ the surface energy, $A_{gb}$ the boundary area, $A_s$ the surface area and $\varepsilon_i$ the total energy of the four relaxed ideal eGBs in the calculated suspended films (see Fig. S1). The defect energy is calculated as $E_N = \sum_{i=1}^{N} e_i - Ne_{coh}$, where $e_i$ and $e_{coh}$ are atomic cohesive energy in the simulation system and in the bulk, respectively [19, 20]. The ideal eGB refers to the boundary in which no atoms have been deleted or inserted. Taking the vacuum energy as the potential reference, the cohesive energy of a single copper atom is $e_{coh} = -3.54eV$ and the defect energy of the single vacancy in bulk is 1.27eV. Both values are consistent with earlier calculations [26].

Then we calculated the defect energy $E_{N-D} = (\sum_{i=1}^{N-D} e_i - (N-D)e_{coh})$ of suspended films, which includes a core-shifted eGB with $D$ atoms deleted. The deleted atoms are added into the copper bulk while inserted atoms are taken from the copper bulk. In case of $D = 0$, the defect system is the same as the relaxed ideal eGB. Comparing the suspended films with relaxed ideal eGBs, the defect energy of the suspended film with a single core-shifted eGB includes the same surface energy, the same boundary energy, but a different eGB energy since any variation of the local surface energy or boundary energy is due to the presence of the triple junction itself. That is $E_{N-D} = E_N^0 + \varepsilon_{cs}$, where $\varepsilon_{cs}$ is the energy of the one core-shifted eGB and 3 ideal eGBs (see Fig. S1). The defect energy variation

$$\Delta E \triangleq E_{N-D} - E_N = \varepsilon_{cs} - \varepsilon_i = (\epsilon_{cse} - \epsilon_{ie}) \cdot l_p \tag{1}$$

is the extensive energy difference between the core-shifted eGB energy $\epsilon_{cse}$ and the corresponding relaxed ideal eGB energy $\epsilon_{ie}$, where $l_p$ is the triple junction length along one period vector.

Figure 2 shows the core shift process for an emergent grain boundary with in-plane angle $\theta$ =13.17° (eGB13.17). We find the boundary core by visualizing only high energy atoms of the calculated system in OVITO [30]. The top panel shows the top view, the middle panel the structure projected along the boundary normal, and the bottom panel the structure projected along the period vector. To shift the boundary core, we delete atoms step by step from the top surface layers of the eGB. The core shift (CS) depth and total deleted atoms (D) are indicated. CS0D0 is the relaxed ideal eGB with zero deleted atoms. After deleting one core atom from each of the top 3 layers of CS0D0, we get CS3D3, whose stacking fault ribbon (SFR) now extends 6 atom layers from the top. After deleting one core atom from each of the top 6 layers of CS3D3, we get CS6D9, whose stacking fault ribbon extends 9 atom layers from the top and the number of the total deleted atoms is 9. Similarly, further deletion yields CS9D18 from CS6D9 and finally CS12D30 from CS9D18. The bottom panel clearly shows the emerging stacking fault ribbon as the number of deleted atoms is increased. The SFR narrows when it meets the bulk boundary core and widens when it meets the free surface. In contrast, the emerging SFR at a ridge eGB becomes narrower when it meets both the bulk boundary core and the free surface (see Fig. S2).

## 3. Energy-depth relation and wedge disclination model

To analyze the resulting core-shifted boundary, we first determine the rotation depth $h$ through which the adjoining grains have been rotated (see AD in Fig. 1c and green dash line in Fig 1d), which is critical to accurately model the energy of systems containing a disclination. Since CSB formation involves grain rotation an incorrect depth gives the wrong coefficients for the quadratic terms for the bulk energy and consequently the wrong shear modulus (see below). Even a systematic offset in the depth gives a different linear term that is related to the interfacial surface energy contribution.

For each [112] CSB at the surface, the number of atomic layers in the stacking fault ribbon (SFR) is designated as $SFRno$ in units of the atomic layer thickness. The number of layers that have been core shifted (CS) and from which atoms have been deleted is $SFRno - 3$, which is clearly seen in Fig. 2. The difference is related to the 3 layers at the transition between the bulk core and SFR core, which are in the SFR but not core shifted. During deletion or insertion of atoms, the corresponding eliminated volume or added volume per period vector length $l_p$ is proportional to the volume variation $(SFRno - \alpha)^2 \cdot l_p$. To determine the appropriate scaling, we determined the numbers of deleted or inserted atoms as a function of the $SFRno$ for four representative eGBs (eGB3.89, eGB13.17, eGB18.73 and eGB26.01). Since the number of deleted atoms $D$ is in principle proportional to the volume, the best proportional fitting in Fig. S3 shows that $\alpha = 1.5$ for each grain boundary and hence the rotation depth is

$$h = (SFRno - 1.5) \cdot a/\sqrt{3} \tag{2}$$

where $a$ is lattice constant and $a/\sqrt{3}$ the layer thickness. This can be understood by recognizing the [111] boundary core can be decomposed into a $1/2$ [112] facet, which involve three atomic layers. When this facet connects with the eGB core, half of the length belongs to the bulk boundary core and the other half belongs to the emergent boundary core. For example, in CS3D3 the SFR depth $SFRno$ and core shifted depth from which the atoms are deleted atoms are 6 and 3, respectively. Hence the rotation depth is 4.5 in unit of atomic layer thickness along [111], which corresponds to the rotation depth $h$ of the wedge disclination associated with CS3D3.

Separately, we can calculate the volume of the wedge-shape disclination whose base is defined by the ABC vertices in Fig. 1c as $h^2 \cdot tan(\varphi/2) \cdot l_p = b \cdot ah^2/8$, where $b$ is number of full dislocations $1/2\,[1\bar{1}0]$ per period vector. This volume divided by the single atom volume is the number of the deleted atoms per period. From this expression for the volume and equation (2), the calculated number of the deleted atoms per period for eGB03.89, eGB13.17 (both have a single core per period or $b = 1$) are $1/6\,(SFRno - 1.5)^2$ and for eGB18.73 and eGB26.01 (that have four cores per period or $b = 4$ ) are $2/3\,(SFRno - 1.5)^2$. The result of this scaling analysis is shown in the top panel of Fig. 3 where all points plotted show an uncertainty of less than one atom. Note this scaling works equally for valley and ridge boundaries, where core atoms are deleted and added, respectively (see Fig. S1). Through the depth analysis of the atomic structure, it is obvious that the volume variation for building core-shifted eGBs can be well described by the wedge disclination.

The calculated defect energies for the four representative eGB structures as a function of the wedge depth are shown in the lower panels of Fig. 3. In each case, we kept the eGB period the same as the bulk boundaries and so we may miss some low energy eGB structures with long periods [31]. Each data point in the figure represents one core-shift [112] eGB structure. We find that the best way to fit the energy points at different depths in Fig. 3 is a parabolic equation [32, 33]. The intensive core-shifted eGB energy can be written as

$$\epsilon_{cse} - \epsilon_{ie} = u \cdot h^2 + v \cdot h + w \tag{3}$$

in units of $J/m$, where $h$ is the rotation depth and $u, v, w$ are coefficients to be determined. The fitting results for valley and ridge eGBs are shown in the graphs in the lower panels in Fig.3 for each of the four representative in-plane angles $\theta$. In all instances, the depth of the disclination wedge is larger for valley eGBs. The parabola shape is determined by coefficient $u$ and the lateral position is determined by the ratio $v/2u$. The vertical position is determined by coefficient $w$. The quadratic term and linear term correspond to the bulk energy and interfacial energies, respectively [32, 33]. The constant corresponds to the energy variation of the linear defect.

To understand the energy-depth dependence of the eGBs in Fig 3, the contributions of the quadratic and linear terms to the defect energy of the core-shifted eGB were analyzed for each of the four representative in-plane angles $\theta$. Figure 4a shows the coefficients of the quadratic term as a function of the corresponding out-of-plane angle $\varphi$ given by $\tan\psi = \tan(\varphi/2)/\sin(\theta/2)$, with $\psi$ =19.5° for each [112] core-shifted eGB. Comparing the quadratic coefficients of $h^2$ at different in-plane angles as shown in Fig. 4a, we found that this coefficient is proportional to $\varphi^2$ and hence the energy contribution is proportional to $\varphi^2 h^2$. This is consistent with finite element calculations in our previous paper [19]. The quadratic term has the form of the linear elastic energy of a straight wedge disclination $G/4\pi(1-\nu) \cdot \varphi^2 h^2$, where $G$ is shear modulus and $\nu$ Poisson's ratio, even though the discrete deletion or insertion of atoms cannot exactly describe a continuum wedge disclination [34-36]. Taking the Voigt average Poisson ratio $\nu = 0.324$ and the average coefficient 6.70 for $\varphi^2 h^2$ based on Fig 4a yields a shear modulus of 56.9 GPa, which is close to the Voigt average value of 54.6 GPa shear modulus in polycrystals [23]. Alternatively, we fitted the energy-depth dependence to $\varphi^2 h^2$ for each valley eGB (see lower panel of Fig 3, black curve), we find that the values of $G/4\pi(1-\nu)$ are 6.28, 6.28, 6.28 and 6.85 corresponding to eGB3.89, eGB13.17, eGB18.73 and eGB26.01, respectively, which is consistent with an average value of 6.70. In summary, the quadratic coefficient is $u = G/(4\pi(1-\nu))$ and a direct measure of the materials elastic properties at their triple junctions.

To analyze the energy contributions to the linear term, we plot in Fig. 4b the coefficient of the linear term (red curve) and the grain boundary energy term (green curve) versus the out-of-plane angle for each valley eGB. The boundary energy term $-(\gamma_{gb[111]} - \gamma_{gb[112]}/\cos(\varphi/2)) \approx -(\gamma_{gb[111]} - \gamma_{gb[112]})$ is a significant driver for CSB formation [28] but from Fig 4b it makes an increasingly smaller contribution to the interfacial energy as both $\theta$ and $\varphi$ increase. In fact, the difference between these curves (blue curve) shows a nearly linear dependence on $\varphi$, with a proportional constant ~0.8, and hence on the wedge width $2h \cdot \tan(\varphi/2)$. As detailed in Tab. S2 for the eGB with lowest energy at different in-plane angles, the wedge width has atomic scale dimensions and is much smaller than the groove width at the surface that is found in experiment and simulation (see below). Therefore, we now consider the effect of the wedge disclination on the surface energy variation and the surface area variation within the linear elastic approximation.

Figure 4c shows the wedge disclination schematically. The period vector $p\mathbf{e}_x$ points out of the paper and Frank vector for the disclination is $\varphi\mathbf{e}_x$. The boundary normal is $n\mathbf{e}_y$. The surface normal is $l\mathbf{e}_z$. Assuming that the total area per period on the left side of the wedge disclination is $\int_{-M}^{0} l_p \cdot dy = A$ and the deformation length during the wedge disclination deformation is $\int_{-M}^{0} \varepsilon(y) dy = h \cdot \tan(\varphi/2)$, the extensive surface energy due to the presence of the wedge disclination deformation is given by:

$$\int_{-M}^{0} \gamma(y) \cdot l_p \cdot dy = \int_{-M}^{0} (\gamma_s + \beta \cdot \varepsilon(y)) \cdot l_p \cdot dy \tag{4}$$

where $M$ is a large distance from the boundary plane so the integration captures all the excess surface energy, and where $\gamma_s$ is the surface energy of a free surface, $\gamma(y)$ the local intensive surface energy, $\varepsilon(y)$ the local uniaxial strain and $\beta$ is the surface stress [37]. Clearly the variation of the excess extensive surface energy due to the formation of the wedge disclination is $-\beta \cdot h \cdot \tan(\varphi/2)$, in contrast to $-\gamma \cdot h \cdot \tan(\varphi/2)$ reported in our previous publication [19]. We find that the surface stress for our four representative eGBs are 0.64, 0.82, 0.89 and 1.15 $J/m^2$, comparable to reported values of surface stress from 0.7 ~ 1.3 $J/m^2$ [37]. Note that the surface stress at ridge eGBs is much smaller than that at valley eGBs. For ridge eGB13.17 and eGB26.01, the surface stress is negative and hence the triple junction is not stable and should spontaneously reorganize. Clearly additional experiment and simulation is needed to understand the behavior of ridge eGB. In summary, the linear coefficient is $v = -(\gamma_{gb[111]} - \gamma_{gb[112]}/\cos\varphi/2) - \beta \cdot \tan(\varphi/2)$ and captures the boundary energy and surface stress contributions to [112] CSB formation.

The constant term $w$ in the parabolic equation Eq. (3) is related to the relaxed ideal eGB, which is sensitive to the precise stacking termination A, B, or C on the surface (see below). Literally, the constant term is equal to the energy difference between the core-shifted eGB with $h = 0$ and $SFRno = 1.5$ and the relaxed ideal eGB. In the special case of an ideal eGB with $SFRno = 1.5$, these two eGBs are the same and the constant term is zero. In most cases, these two eGBs are different and hence the constant term $w$ is not zero. A more detailed analysis will be provided in a future publication.

The above analysis shows that the elastic energy stored in the bulk around the triple junction of [112] CSBs is well described by the straight wedge disclination model at the free surface, even though the wedge width is as short or shorter than an atomic distance (see Tab. S2). In addition, the interfacial energy of the eGBs due to the variation of the surface area and the surface stress can also be described by the wedge disclination model. Furthermore, the relaxed ideal eGB without any deleted atoms still has nonzero SFR depth, such as CS0D0 in Fig. 2, where the SFR is three layers deep. The ideal eGB is included in the above analysis and is well described as a wedge disclination with a depth of an atomic distance.

## 4. Multiple core eGB

Considering the differences and complexity of eGBs with multiple cores per period vector and the necessity to demonstrate that core-shifted boundaries can describe the structure found in experiment, we show in Fig. 5 the atomic structure of eGB26.01. A systematic structure calculation showed that the lowest energy structure (wedge depth 1.04 nm in Fig. 3h) involves the deletion of 16 atoms per period, consistent with our experimental measurement of the groove volume [19], with the number of deleted atoms removed sequentially from each layer (starting from the topmost layer) is: 6, 4, 4, 2, respectively. The wedge depth described in Fig. 3 is the average value over the 4 boundary cores.

The surface topography and subsurface structure for eGB26.01 are shown in Fig. 5a and Fig. 5b-c, respectively. At the atomic scale on the surface, the period vector decomposes as [3,2,3,2], consistent with experiment [19]. On the nanoscale level, the surface is deformed to create a groove of width ~ 5 nm along the boundary plane, slightly wider than our experimental value [19]. The calculation also accurately captured the corrugation nonequivalence of the four dislocations in each period. Except for the 7 atoms highlighted by the two ellipses in Fig 5a located at a depth of ~ 30 pm, the groove depth is ~ 50 pm, comparable with experimental value ~ 60 pm (see Fig. S3a). The highlighted atoms at reduced depths are associated with dislocations 2 and 4 in Fig. 5a and the $SFRno = 6$ as seen in Fig. 5b-c. The deletion of atoms results in an emergent boundary that is under tensile stress [2], consistent with the hydrostatic stress distribution shown in Fig. S3b [38].

## 5. Conclusion

We have shown that [112] core-shifted boundaries can be well described as a straight wedge disclination at the free surface. The subsurface geometry found is consistent with our previous geometrical analysis of experimental data. The determination of the SFR depth and in turn the depth of the disclination wedge is supported by our deleted atom and volume analysis. The results derived from an analysis of the quadratic and linear terms of the energy-depth relationship in Eq. 3 yield elastic parameters in bulk and surface stresses that are consistent with reported values.

In our calculation for each representative eGB, the boundary core of the lowest-energy structures at different depths had a local [112] tilt axis and their stacking fault ribbons were always in the close-packed plane, consistent with our geometrical analysis of experimental data [19]. For films of a fixed thickness, the depth of the stacking fault ribbon and the bulk GB core vary every three atomic layers. In building the disclination wedge via the core-shift method (see Fig 2) when the depth of the stacking fault ribbon increases by three atomic layers, the bulk boundary core decreases three atomic layers. In a future paper we will discuss the behavior when additional single layers are added such as an A or B terminated slab, in contrast to the C terminated slab discussed here. Such effects are important whenever single height atomic steps cross the boundary plane, as is common experimentally.

The core shift method introduced here can be used as the starting point for research on other low energy structures beyond [112] wedge disclination. In this manner, it is possible to screen for the lowest energy structures that are consistent with experimental results, for example by fixing the CS depth and varying the out-of-plane angles (and hence the local tilt angle) through the insertion of jogs along the SFR. However, not all core-shift angle can be built. On one hand, a minimum depth is necessary to build a certain out-of-plane rotation due to the discrete nature of the atoms that are deleted or added. For example, adding one additional jog into [112] core-shifted eGB shifts the composite rotation axis a certain angle away from [112]. On the other hand, the CS depth is determined by the interplay between interfacial energies and elastic energy. Due to this complexity, the corresponding results are not included in the present publication. On this basis, however, the core shift method can also be used to describe low energy structures at emergent metastable boundaries, to identify low energy structures in clusters and nanoparticles with dislocations, or to build subsurface metastable structures, and even to carry out structure searches at triple junctions in bulk.

J.J.B, X.Z. and I.D. acknowledge support from Science Foundation Ireland grants (12/RC/2278 and 16/IA/4462) and thank Trinity Centre for High Performance for providing the computing resource.

**Figure 1 Boundary geometry**
(a) the median lattice adhering to the $xyz$ coordination system. (b) bicrystals and their own coordination system are rotated $\pm\theta/2$ along axis z or [111] of the median lattice. (c) out-of-plane rotation $\varphi$, to shift the composite rotation axis $l$. (d) [112] core-shifted boundary at top surface implied from experiment [19]. (e) core shift method illustrated with a pure edge dislocation in simple cubic lattice [22]. The grey atoms show the inserted plane. Deleting the core atom or adding one atom below the core ⊥ shifts the dislocation line.

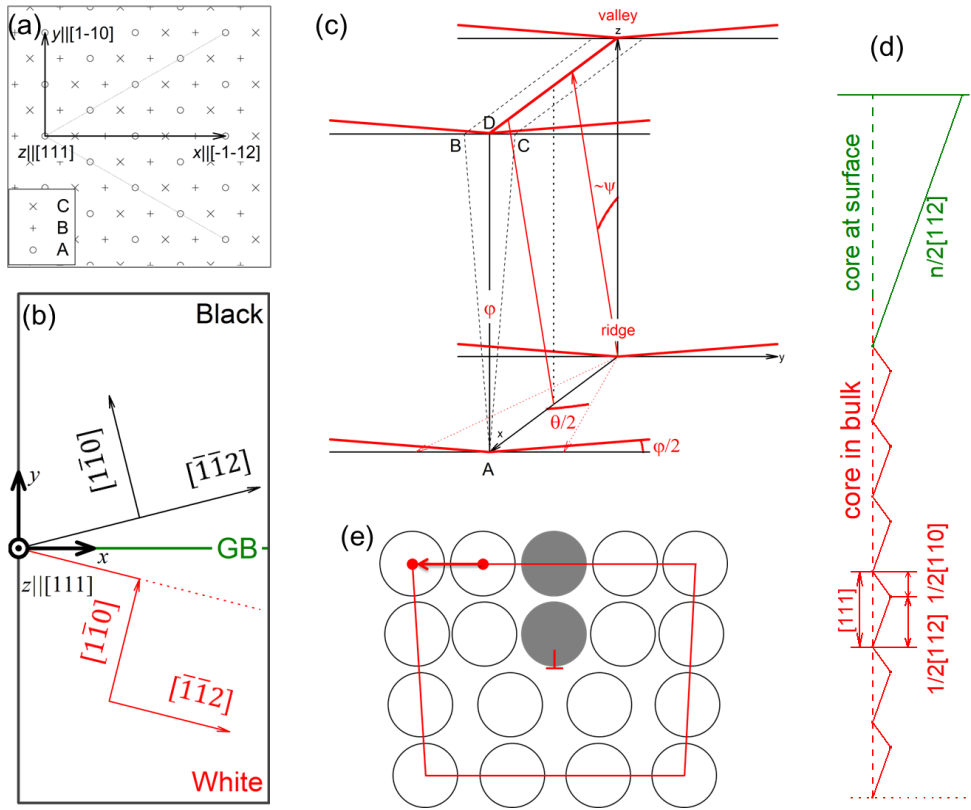

**Figure 2 Core-shift boundary with in-plane angle 13.17°**
To shift the boundary core, we delete atoms at eGB in a step-by-step manner from the top surface. Core shift (CS) depth and total deleted atoms (D) are indicated. CS0D0 is the relaxed ideal eGB. After deleting one core atom each layer from the top 3 layers of CS0D0, we get CS3D3, whose stacking fault ribbon goes through 6 atoms layer from the top. After deleting one core atom each layer from the top 6 layers from CS3D3, we get CS6D9, whose stacking fault ribbon goes through 9 atoms layer from the top.

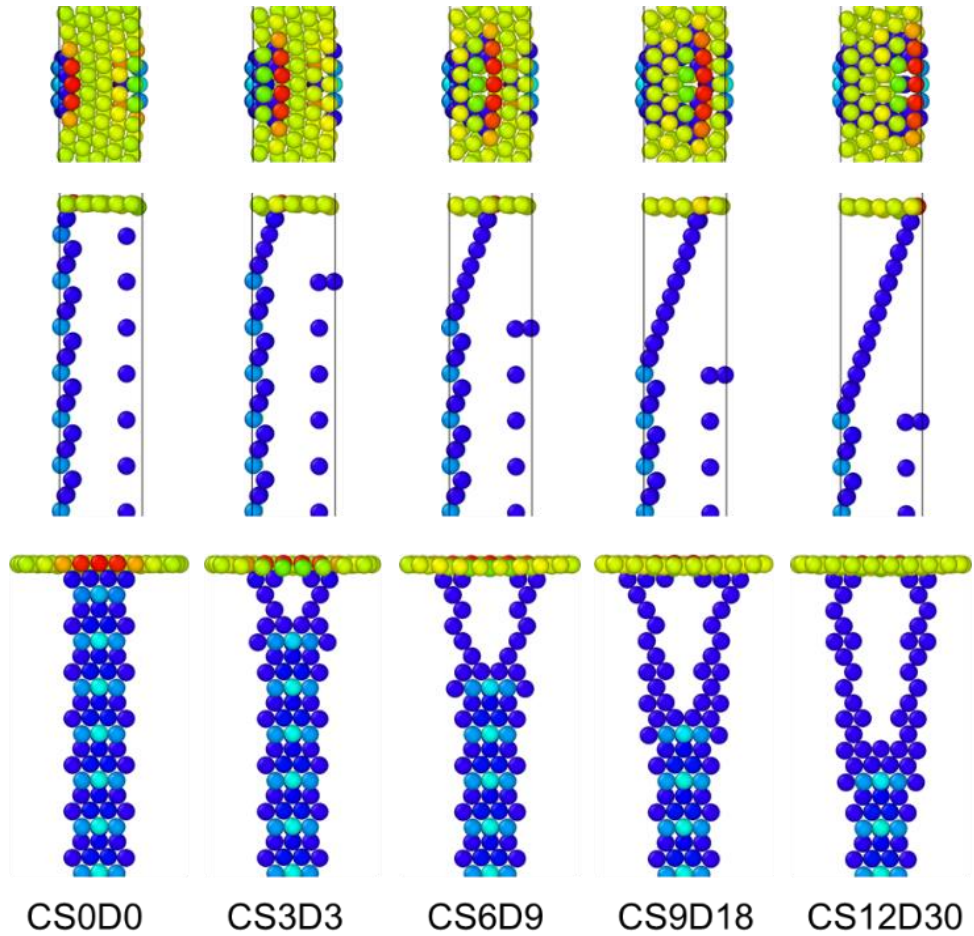

**Figure 3 Number of deleted atoms and core-shifted eGB energy**

Upper panels show the relationship between number of deleted atoms and effective depth from SFR thickness for representative eGBs with in-plane angle 3.89, 13.17, 18.73, 26.01. (b, d, f, h) Lower panel show dependence of eGB energies on wedge depth. Valley and ridge boundaries are shown as black and red curves, respectively. The fitting parameters to Eq. 3 are shown for each in-plane angle. The suspended film thickness used is shown - for the low energy boundary with in-plane angle 3.89, a thick suspended film with 200 ABC layer stacking or 600 total layers and nearly 10 million atoms was used.

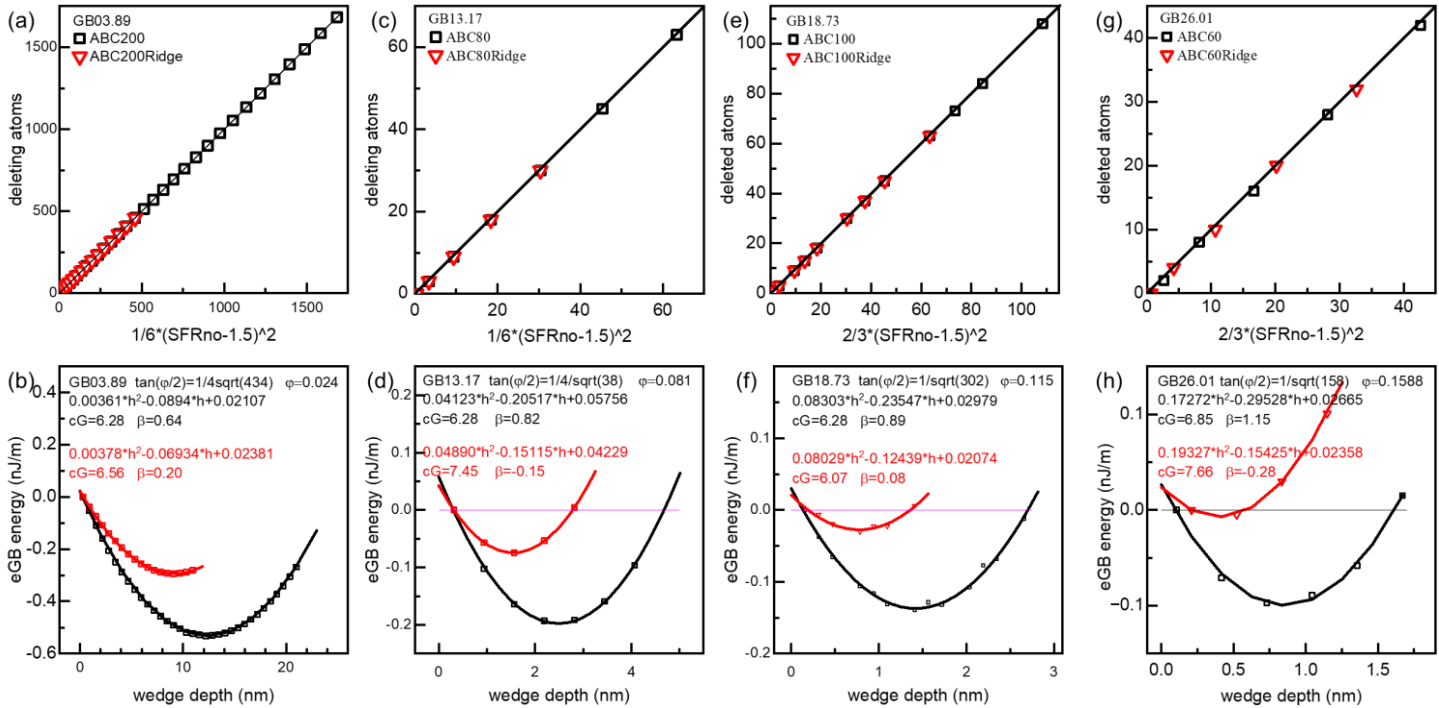

**Figure 4 Analysis of the quadratic term & linear terms**

(a) Dependence of the coefficient $u$ of $h^2$ on out-of-plane angle. Data points are for eGB3.89, eGB13.17, eGB18.73 and eGB26.01. This plot demonstrates that the quadratic term of the defect energy is proportional to $\varphi^2 h^2$. (b) Plot of the linear term coefficient $v$, boundary term, the difference between previous two terms versus out-of-plane angle. The proportional fitting of the difference with constant ~ 0.8 is also shown. (c) Schematic show of one straight wedge disclination at the free surface. The period vector $pe_x$ points out of the paper and Frank vector for the disclination is $\varphi e_x$. The boundary normal is $ne_y$. The surface normal is $le_z$.

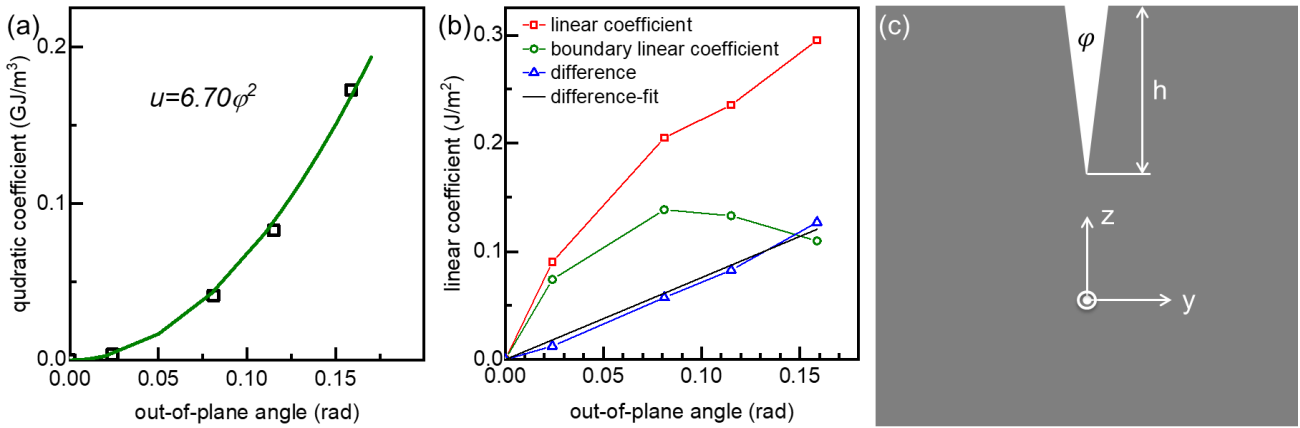

**Figure 5 Subsurface eGB26.01 structure**

(a) Surface structure of the lowest energy structure. The length along the period vector is two times of the period. The four dislocations in one period are labelled. (b) Boundary cores projected onto the boundary plane. Atoms are coloured by its energy and atoms of lower energy are deleted. (c) visualizing four boundary cores projected along the period vector.

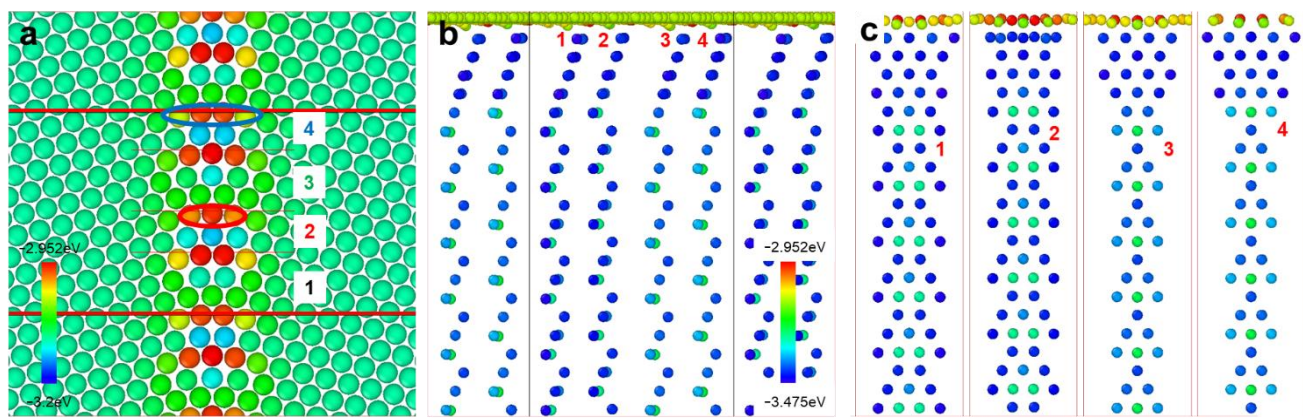

**References**
[1] A.P. Sutton, R.W. Balluffi, Interfaces in Crystalline Materials, OUP Oxford2006.
[2] W.D. Nix, B.M. Clemens, Crystallite coalescence: A mechanism for intrinsic tensile stresses in thin films, J Mater Res 14(8) (1999) 3467-3473.
[3] E. Chason, A kinetic analysis of residual stress evolution in polycrystalline thin films, Thin Solid Films 526 (2012) 1-14.
[4] C. Friesen, C.V. Thompson, Comment on "Compressive stress in polycrystalline Volmer-Weber films", Phys Rev Lett 95(22) (2005) 229601; author reply 229602.
[5] C. Friesen, C.V. Thompson, Correlation of stress and atomic-scale surface roughness evolution during intermittent homoepitaxial growth of (111)-oriented Ag and Cu, Phys Rev Lett 93(5) (2004) 056104.
[6] C. Friesen, C.V. Thompson, Reversible stress relaxation during precoalescence interruptions of volmer-weber thin film growth, Phys Rev Lett 89(12) (2002) 126103.
[7] J. Leib, R. Monig, C.V. Thompson, Direct evidence for effects of grain structure on reversible compressive deposition stresses in polycrystalline gold films, Phys Rev Lett 102(25) (2009) 256101.
[8] R. Koch, D.Z. Hu, A.K. Das, Comment on "Compressive stress in polycrystalline Volmer-Weber films" - Reply, Phys Rev Lett 95(22) (2005).
[9] R. Koch, D. Hu, A.K. Das, Compressive stress in polycrystalline volmer-weber films, Phys Rev Lett 94(14) (2005) 146101.
[10] C.W. Pao, S.M. Foiles, E.B. Webb, D.J. Srolovitz, J.A. Floro, Thin film compressive stresses due to adatom insertion into grain boundaries, Phys Rev Lett 99(3) (2007) 036102.
[11] R.G. Mariano, M. Kang, O.J. Wahab, I.J. McPherson, J.A. Rabinowitz, P.R. Unwin, M.W. Kanan, Microstructural origin of locally enhanced CO2 electroreduction activity on gold, Nat Mater 20(7) (2021) 1000.
[12] R.G. Mariano, K. McKelvey, H.S. White, M.W. Kanan, Selective increase in CO2 electroreduction activity at grain-boundary surface terminations, Science 358(6367) (2017) 1187-1192.
[13] X. Zhao, T.K. Gunji, F. Lv, B.L. Huang, R. Ding, J.G. Liu, M.C. Luo, Z.G. Zou, S.J. Guo, Direct Observation of Heterogeneous Surface Reactivity and Reconstruction on Terminations of Grain Boundaries of Platinum, Acs Mater Lett 3(5) (2021) 622-629.
[14] D. An, T.A. Griffiths, P. Konijnenberg, S. Mandal, Z. Wang, S. Zaefferer, Correlating the five parameter grain boundary character distribution and the intergranular corrosion behaviour of a stainless steel using 3D orientation microscopy based on mechanical polishing serial sectioning, Acta Mater 156 (2018) 297-309.

[15] M. Bettayeb, V. Maurice, L.H. Klein, L. Lapeire, K. Verbeken, P. Marcus, Nanoscale Intergranular Corrosion and Relation with Grain Boundary Character as Studied In Situ on Copper, J Electrochem Soc 165(11) (2018) C835-C841.
[16] E. Martinez-Lombardia, L. Lapeire, V. Maurice, I. De Graeve, K. Verbeken, L.H. Klein, L.A.I. Kestens, P. Marcus, H. Terryn, In situ scanning tunneling microscopy study of the intergranular corrosion of copper., Electrochem Commun 41 (2014) 1-4.
[17] C. Wang, K. Du, K. Song, X. Ye, L. Qi, S. He, D. Tang, N. Lu, H. Jin, F. Li, H. Ye, Size-Dependent Grain-Boundary Structure with Improved Conductive and Mechanical Stabilities in Sub-10-nm Gold Crystals, Phys Rev Lett 120(18) (2018) 186102.
[18] Y.-Y. Zhang, H. Xie, L.-Z. Liu, H.-J. Jin, Surface Triple Junctions Govern the Strength of a Nanoscale Solid, Phys Rev Lett 126(23) (2021) 235501.
[19] X. Zhang, M. Wang, H. Wang, M. Upmanyu, J.J. Boland, Restructuring of emergent grain boundaries at free surfaces – An interplay between core stabilization and elastic stress generation, Acta Mater 242 (2023) 118432.
[20] X. Zhang, J. Han, J.J. Plombon, A.P. Sutton, D.J. Srolovitz, J.J. Boland, Nanocrystalline copper films are never flat, Science 357(6349) (2017) 397-400.
[21] J. Schiotz, K.W. Jacobsen, Nanocrystalline metals: Roughness in flatland, Nat Mater 16(11) (2017) 1059-1060.
[22] G.R. Love, Dislocation pipe diffusion, Acta Metallurgica 12(6) (1964) 731-737.
[23] J.P. Hirth, J. Lothe, Theory of Dislocations, McGraw-Hill1982.
[24] N.F. Mott, The Mechanical Properties of Metals, P Phys Soc Lond B 64(381) (1951) 729-&.
[25] S. Plimpton, Fast Parallel Algorithms for Short-Range Molecular-Dynamics, J Comput Phys 117(1) (1995) 1-19.
[26] Y. Mishin, M.J. Mehl, D.A. Papaconstantopoulos, A.F. Voter, J.D. Kress, Structural stability and lattice defects in copper: Ab initio, tight-binding, and embedded-atom calculations, Phys Rev B 63(22) (2001) 224106.
[27] M.A. Tschopp, D.L. McDowell, Asymmetric tilt grain boundary structure and energy in copper and aluminium, Philos Mag 87(25) (2007) 3871-3892.
[28] X. Zhang, J.J. Boland, Universal preference for low energy core-shifted grain boundaries at the surfaces of fcc metals, submitted (2023).
[29] P. Muller, A. Saul, Elastic effects on surface physics, Surf Sci Rep 54(5-8) (2004) 157-258.
[30] A. Stukowski, Visualization and analysis of atomistic simulation data with OVITO-the Open Visualization Tool, Model Simul Mater Sc 18(1) (2010) 015012.
[31] D.L. Olmsted, D. Buta, A. Adland, S.M. Foiles, M. Asta, A. Karma, Dislocation-pairing transitions in hot grain boundaries, Phys Rev Lett 106(4) (2011) 046101.
[32] T. Radetic, F. Lancon, U. Dahmen, Chevron defect at the intersection of grain boundaries with free surfaces in Au, Phys Rev Lett 89(8) (2002) 085502.
[33] T. Radetic, C. Ophus, D.L. Olmsted, M. Asta, U. Dahmen, Mechanism and dynamics of shrinking island grains in mazed bicrystal thin films of Au, Acta Mater 60(20) (2012) 7051-7063.
[34] A.E. Romanov, A.L. Kolesnikova, Application of disclination concept to solid structures, Prog Mater Sci 54(6) (2009) 740-769.
[35] A.E. Romanov, V.I. Vladimirov, Straight Wedge Disclination near a Free-Surface, Phys Status Solidi A 59(2) (1980) K159-K163.
[36] A.E. Romanov, A.L. Kolesnikova, Elasticity Boundary-Value Problems for Straight Wedge Disclinations. A Review on Methods and Results, Reviews on Advanced Materials and Technologies 3 (2021) 55-95.
[37] C.W. Pao, D.J. Srolovitz, C.V. Thompson, Effects of surface defects on surface stress of Cu(001) and Cu(111), Phys Rev B 74(15) (2006) 155437.
[38] V. Turlo, T.J. Rupert, Linear Complexions: Metastable Phase Formation and Coexistence at Dislocations, Phys Rev Lett 122(12) (2019) 126102.